# A New Biometric Template Protection using Random Orthonormal Projection and Fuzzy Commitment

Thi Ai Thao Nguyen, Tran Khanh Dang and Dinh Thanh Nguyen

Faculty of Computer Science and Engineering
Ho Chi Minh University of Technology, VNU-HCMU, Vietnam
{thaonguyen, khanh, dinhthanh}@hcmut.edu.vn

**Abstract.** Biometric template protection is one of most essential parts in putting a biometric-based authentication system into practice. There have been many researches proposing different solutions to secure biometric templates of users. They can be categorized into two approaches: feature transformation and biometric cryptosystem. However, no one single template protection approach can satisfy all the requirements of a secure biometric-based authentication system. In this work, we will propose a novel hybrid biometric template protection which takes benefits of both approaches while preventing their limitations. The experiments demonstrate that the performance of the system can be maintained with the support of a new random orthonormal project technique, which reduces the computational complexity while preserving the accuracy. Meanwhile, the security of biometric templates is guaranteed by employing fuzzy commitment protocol.

## 1 Introduction

Biometric based authentication systems have been applied more and more in our daily life. Nowadays, we can access to our personal devices by providing to sensors our fingerprint, face, voice, iris,… We also do the same thing in order to access to thousands of online services. There is no need to remember or carry anything like password, token,… to prove ourselves to systems. It is obviously so convenience. However, benefits always come along with challenges. These challenges are related to the security of biometrics. As we all know, human has a limited number of biometric traits; therefore, we cannot change our biometrics frequently like password once we suspect that the templates are revealed [1]. Moreover, the fact that people may register thousands of online services makes them use the same biometrics to sign in some services. That leads to the cross-matching attacks when attackers follow user's biometric template cross the online services in order to track their activities. Another concern relates to the natural set-backs of biometrics. The fact that biometrics reflects a specific individual means it contains sensitive information (e.g. medical conditions) which users do not want attacker or even the server storing users' authentication data to discover. Last but not least, the network security needs to be discussed when user's private information is transmitted over insecure network [2]. As a result, biometric template protection is apparently indispensable part in every biometric system applications. To solve the security issues, the commonly approach is to store the transformed biometric template instead of the original one. According to [1], there are three requirements that a biometric template protection technique should possess.

1. Cancelability (Revocability + Diversity): it should be straightforward to revoke a compromised template and reissue a new one based on the same biometric data. In addition, the scheme should not generate the same transformed templates of an individual for different applications.
2. Security: An original biometric template must be computationally hard to recover from the secure template. This property guarantees that an adversary does not have the ability to create a physical spoof of the biometric trait from a stolen template.
3. Performance: the biometric template protection scheme should not degrade the recognition performance of the biometric system. This requirement can refer to the discriminability of the original biometric template which should be preserved after transformed. It also means transformed templates from the same user should have high similarity in the transformed space. And the ones from different users should be dissimilar after transformed.

To meet as many requirements as possible, many techniques have been proposed which can be classified into two main approaches: the feature transform approach and the biometric cryptosystem approach.

In the first approach, biometric templates are transformed using a function defined by a user-specific factor such as a key, a password, or a random string… The goal of this approach is to provide diversity and unlinkability by using different transforming functions for different applications involving the same set of users. The intrinsic strength of biometric characteristics should not be reduced applying transforms (represented by FAR – False Accept Rate) while on the other hand transforms should be tolerant to intra-class variation (represented by FRR – False Reject Rate). According to [3], two main categories of this approach are distinguished:

1. Salting: This transformation is invertible to large extent, therefore, the factor (also called an authentication key) has to be kept in secret. Thanks to the key, the cancelability requirement is guaranteed, and it also results in low false accept rate. However, the main drawback is also lied on this key. If the key is compromised, the original biometric template can be revealed. Some examples of salting approach have been proposed in previous works [4-6].
2. Non-invertible transform: This approach applies a noninvertible transformation function to protect the original biometric template. Noninvertible transformation refers to a one-way function which is "easy to compute" but "hard to invert". The key to generate this function can be public. Even if an adversary knows the key and the transformed template, it is computationally hard to recover the original biometric template. For this point, this approach provides better security than the salting. The cancelability is also easily achieved by changing the key which generates the transformation function. However, the main drawback is the tradeoff between discriminability and non-invertibility of the transformation function. It is difficult to design a transformation function which satisfies both the discriminability and non-invertibility properties as the same time. In addition, each function is usually suitable for each kind of biometric feature. In [7, 8], the authors presented some proposals for generating cancellable fingerprint templates. In [9], the robust hash function was designed to transform face templates…

The biometric cryptosystems were originally developed for securing a cryptographic key using biometric data or directly generating a cryptographic key from biometric data. However, it has also been used in biometric template protection. In this approach, some helper data which is public is stored in the database. While the helper data does not reveal any significant information about the original biometric template, it is needed during matching to extract a cryptographic key from the query biometric data. According to [3], two main categories of this approach are distinguished: key-binding and key generation

1. Key-binding: In this approach, the original biometric template is bound with a key within a cryptographic framework. The result of this combination is stored in the database as helper data. This helper data does not reveal much information about key or the biometric template. However, the huge setback of this approach is the lack of the cancelability property because it is clearly not designed for this. In addition, the matching process depends on the error correction scheme applied. This can possibly lead to a reduction in the matching accuracy. So many techniques belonging to this approach have been proposed, e.g. fuzzy commitment scheme in [10], fuzzy vault scheme in [11] and its enhanced version for fingerprint, face and iris in [12-14], shielding functions in [15],…
2. Key generation: In this approach, the authentication key is directly generated from biometric template. Therefore, there is no need to be concerned about the key security. However, it usually suffer from low discriminability which can be assessed in term of *key stability* and *key entropy*. Key stability refers to the extent to which the key generated from the biometric data is repeatable, and key entropy relates to the number of possible keys that can be generated. For example, if a scheme generates the same key regardless of the input template, it has high key stability but zero key entropy leading to high false accept rate. Otherwise, if a scheme generates different keys for different templates of the same user, the scheme has high entropy but no stability leading to high false reject rate. Therefore, the limitation of this approach is it is difficult to generate key with high stability and entropy. Moreover, the key generation seems to lose the cancelability property. Some techniques belonging to this approach have been proposed in [16-20] for fingerprint, face,…

From this view, it can be seen that no one single template protection approach can simultaneously satisfy three both requirements (cancelability, security, and performance). On this account, some recent studies tend to integrate the advantages of both approaches while avoiding their limitations. Hybrid approach is the combination two or more methods to create a single template protection scheme. Very recently, in 2018, the combination of secure sketch and ANN (Artificial Neural Network) was proposed [21]. The ANN with high noisy tolerance capacity can not only enhance the recognition by learning the distinct features, but also assure the revocable and non-invertible properties for the transformed template. In addition, the secure sketch's construction can reduce the false rejection rate significantly due to its error correction ability. The fuzzy vault was combined with periodic function based transformation in [22], or with the non-invertible transformation to conduct a secure online authentication in [23]. The homomorphic cryptosystem was employed in fuzzy commitment scheme to achieve the blind authentication in [24]. Another combination approach was introduced in [25]. In this work, we are going to integrate the random orthonormal project and ideal of fuzzy commitment to guarantee the security for user's biometric template.

## 2  Background

### 2.1  Random Orthonormal Projection

Random Orthonormal Projection (ROP) is a technique that utilizes an orthonormal matrix to project a set of points into other space while preserving the distances between points. It was firstly presented as a secure transform for biometric templates in [4], and its remarkable feature is the ability to meet the revocability requirement. When considered as a standalone biometric template protection, it can be classified into a salting approach by the categorization of Jain [3]. As discussed in Section I, the salting approach has low security because the secure template can be inverted if the key is revealed. Therefore, in [26] the authors proposed an additional module for ROP, which is considered as quantization module (Discriminability Preserving), to make it noninvertible. In this work, we try another way to ensure the secure template stored in database cannot be inverted.

The main idea of random projection is to create $k$ orthonormal vectors size of $l$ ($l$ is also the size of the feature vector extracted from an original biometric template and usually $l = k$). To generate these $k$ orthonormal vectors, $k$ pseudo random vectors are created first and then an orthonormalization process like Gram-Schmidt process is employed to transform these random vectors into the orthonormal ones. This process succeeds if and only if the input vectors are linearly independent. However, this condition cannot always be guaranteed when a set of random vectors are generated. In summary, generating an orthonormal matrix of size $k \times k$ from Gram-Schmidt process may be a critical problem when applied on constraint computationally devices like PDA and handheld devices

Another method to effectively deliver orthonormal matrix was introduced in [27]. The main idea of this new method is based on the fact that a small size orthonormal matrix can be generated without applying Gram-Schmidt process. This matrix is presented as below:

$$I_\theta = \begin{bmatrix} \cos\theta & \sin\theta \\ -\sin\theta & \cos\theta \end{bmatrix} (\forall \theta \in [0, 2\pi])$$

In the same way, we can create an orthonormal matrix $A$ of size $2n \times 2n$ owns a diagonal which is a set of $n$ orthonormal matrices of size $2 \times 2$. The other entries of $A$ are zeros. We present the example of matrix $A$ of size $2n \times 2n$ as shown in the formula below where the values $\{\theta_1, \theta_2, \ldots, \theta_n\}$ are the random numbers in the range $[0: 2\pi]$

$$A = \begin{bmatrix} I_{\theta 1} & 0 & \cdots & 0 \\ 0 & I_{\theta 2} & \cdots & 0 \\ \vdots & \vdots & \ddots & \vdots \\ 0 & 0 & \cdots & I_{\theta n} \end{bmatrix}$$

$$A = \begin{bmatrix} \cos\theta_1 & \sin\theta_1 & \ldots & \ldots & 0 \\ -\sin\theta_1 & \cos\theta_1 & \ldots & \ldots & 0 \\ \vdots & \vdots & \ddots & \ddots & \vdots \\ \vdots & \vdots & \ddots & \ddots & \vdots \\ 0 & 0 & \cdots & \cos\theta_n & \sin\theta_n \\ 0 & 0 & \cdots & -\sin\theta_n & \cos\theta_n \end{bmatrix} \quad (\forall \theta_i \in [0, 2\pi])$$

In result, given the biometric feature vector x of size 2n, orthonormal random matrix A of size $2n \times 2n$, random vector b of size 2n, we have the transformation

$$y = Ax + b$$

Whenever, users have a thought that their secure biometric templates are compromised, these transformed templates can be replaced by the new ones which can be easily constructed by choosing a new set of values $\{\theta_1, \theta_2, \ldots, \theta_n\}$ and vector b.

By using this technique to produce the orthonormal matrix, there is no need for a complex process such as Gram-Schmidt. Beside its effectiveness in computational complexity, it can also improve the security while guaranteeing intra-class variation.

### 2.2 Fuzzy Commitment

Fuzzy commitment scheme belongs to the key-binding approach in biometric cryptosystem [10]. It is the combination between Error Correcting Codes (ECC) and cryptography. To understand how fuzzy commitment scheme works, we have to learn about ECC. Formally speaking, ECC plays a central role in the fuzzy commitment scheme. ECC checks and corrects the corrupted messages if they contains a certain number of errors which this ECC can afford to check.

In fuzzy commitment scheme, a biometric data is treated as a corrupted codeword. During enrollment phase, a user registers biometric template $B_T$, randomly chooses a key $K$ (considered as a codeword). Helper Data Extraction module computes the helper data $HD$ of biometric template $B_T$ and key $K$. During authentication phase, a biometric query $B$ ($B$ and $B_T$ can be quite different because of noise) and the helper data $HD$ are put into the Recover module to extract the key $K'$. If the difference between $B$ and $B_T$ is smaller than the error correction capability of the ECC employed in this fuzzy commitment scheme, the Recover module can recover exactly the same key. The extracted key $K'$, then, is checked with the enrollment key $K$ to decide whether the biometric query is valid or not. This process is demonstrated in Fig.1.

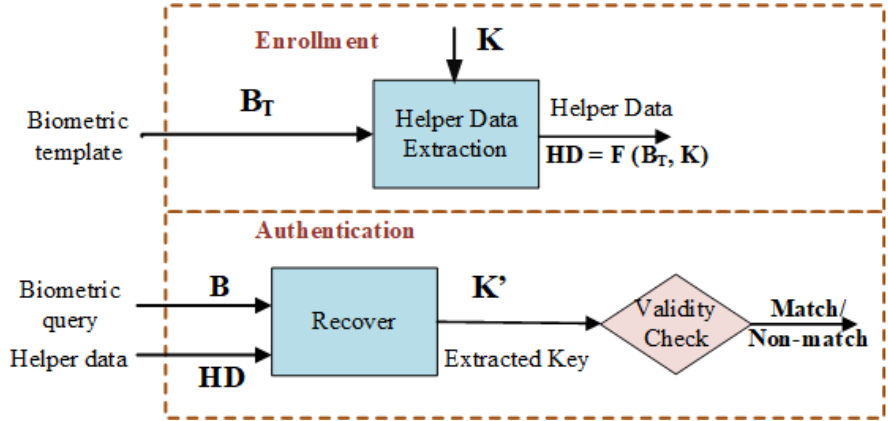

**Fig. 1.** Fuzzy commitment scheme.

## 3 Proposal scheme

The main idea of this hybrid scheme is to secure a non-invertible cryptographic key from cancelable templates. The proposal assumes that the extracted feature of a biometric template can be presented by a vector in continuous domain. Fig. 2 depicts the main idea of the proposal scheme.

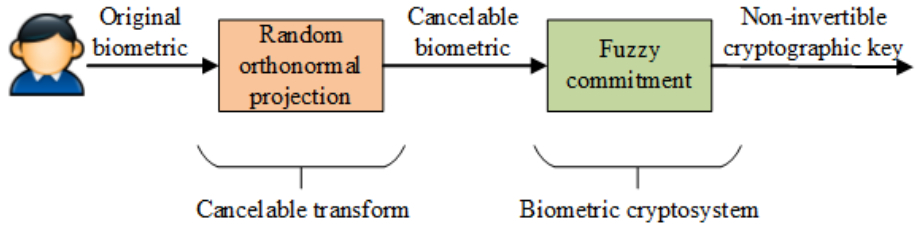

**Fig. 2.** The main idea of proposal scheme.

### 3.1 Enrollment Phase

In the enrollment phase, the user creates a random number $K_M$ which is used to generate a random orthonormal matrix $M$ (as described in section 2.1). This $K_M$ is stored on the user's device to regenerate the matrix $M$ for authentication phase. If user wants to change the transformed template, he/she just has to create new value of number $K_M$. The feature extraction module extracts a feature vector $B$ from the input biometric data. Then, B is combined with matrix M to produce the cancelable version $B_{TC}$. The user also creates a random key $K$. The fuzzy commitment scheme is applied with the inputs (cancelable biometric $B_{TC}$ and the random key $K$). The output of this process is the helper data $HD$ which is stored in the database after that. User also stores the hash version of *K* into the database for the authentication phase. The enrollment phase is demonstrated in Fig. 3.

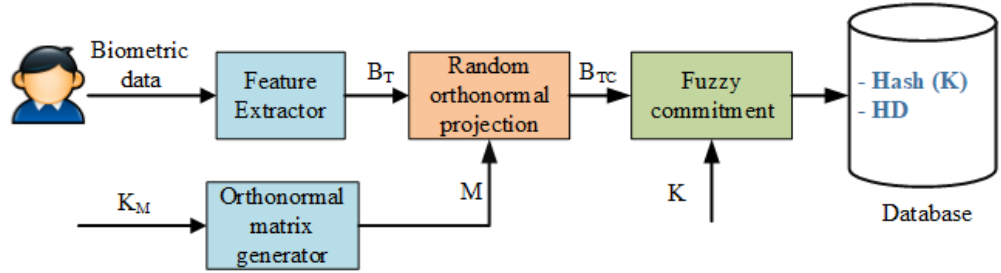

**Fig. 3.** Enrollment phase.

### 3.2   Authentication Phase

In this phase, user provides his/her query biometric data to feature extractor module. It results in the feature vector *B* (note that with the same user, *B* and $B_T$ cannot be the same because of noise). Meanwhile, user retrieves the $K_M$ in his/her device, and put it into the orthonormal matrix generator module to regenerate the matrix *M*. Then, B and M are put into the random orthonormal projection module to create another version of B, which is $B_C$. Since the orthonormal projection provides the discriminability, if $B_T$ and *B* are extracted from the same person, their corresponding cancelable versions $B_{TC}$ and $B_C$ must be similar. In this case, the fuzzy commitment is absolutely able to recover the same key, which the user provided in enrollment phase, from the cancelable biometric $B_C$ and the helper data *HD* retrieved from the database. We call the result of this process *K'*. The hash version of *K'* is compared with *hash(K)* stored in the database. If they are matched, the user is authenticated; otherwise, the authentication process rejects the access of this user. In this scheme, only *hash(K)* and *HD* are stored. These two parameters cannot recover or invert to the cancelable biometric data or even the original biometric data. Therefore, we can affirm that this combination provides the non-invertible transform. The whole process of authentication phase is illustrated in Fig. 4.

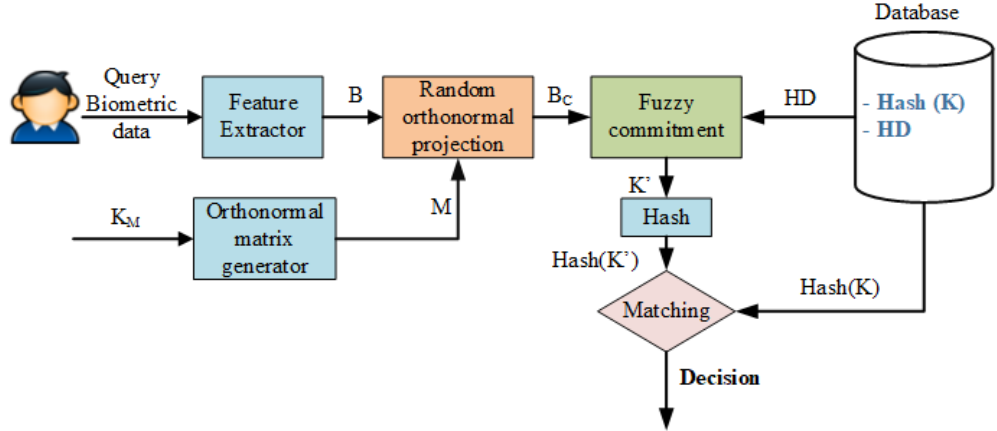

**Fig. 4.** Authentication phase.

## 4     Experimental Result

In this proposal, we employ the set of facial images of many people to get the experimental result. PCA is applied to extract feature vectors from user's faces. Each feature vector has size of 200, and the value of each element is in range [0,1]. PCA is trained under the training data set containing 50 images of 10 South East Asians, 22 Middle and West Asians, and 18 Europeans. The testing data set includes 153 people; each has 20 different facial expression. The first image of each user is registered in the enrollment phase, and the others are used to be tested. The accuracy of the biometric authentication system is evaluated through these error rates: FAR, FRR, EER. The meanings of these rates are explained below:

- FAR, also known as False Acceptance Rate, accepts an entrance when a visitor is invalid. This shows probability of the imposter logging in and succeeding.
- FRR, also known as False Reject Rate, rejects an entrance when a visitor is valid. This shows probability of the visitor logging in & getting rejected
- EER, also known as Equal Error Rate, is intersection of FAR & FRR, at which FAR equals FRR.

In any biometric based authentication system, we need to decide the parameter called threshold. If match score is greater than this threshold, the person who sends the request will be authenticated, otherwise, the system will refuse the request. The value of threshold t is calculated by the formula

$$t_i = 0.1 + 0.01 \times i, \quad for\ i \in \mathbb{N}, and\ i \in [0, 49]$$

For each value of threshold, the values of FAR and FRR are calculated by statistics.

Fig. 5 shows the recognition accuracy results in term of FAR and FRR in cases of no biometric template protection, and Fig. 6 shows the result in cases of applying our hybrid scheme.

In the first case illustrated in Fig. 5, the FRR and FAR intersect at the threshold $t \approx 0.22$. At this intersection, the error rate is about 9%.

In the hybrid scheme, integrating random orthonormal projection and fuzzy commitment, which is demonstrated in Fig. 6, the intersection of FRR and FAR (also known as EER) all values at 9%. This figure proves the proposed hybrid scheme delivers a positive result with the probability of correct recognition of around 91% (the threshold value also depends on the quantizing value; however, the result stays the same; in this experimental result, the quantizing value stands at 200). Hence, it is pertinent that the recognition performance of our hybrid scheme is competitive with the non-template protection ones. In summary, the combination of orthonormal matrix and fuzzy commitment in biometric-based authentication system is absolutely feasible and can be put into practice.

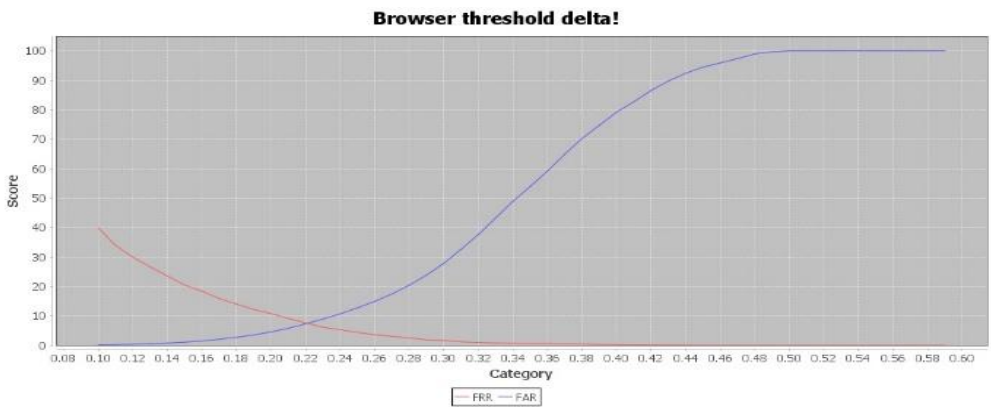

**Fig. 5.** FAR and FRR with no template protection.

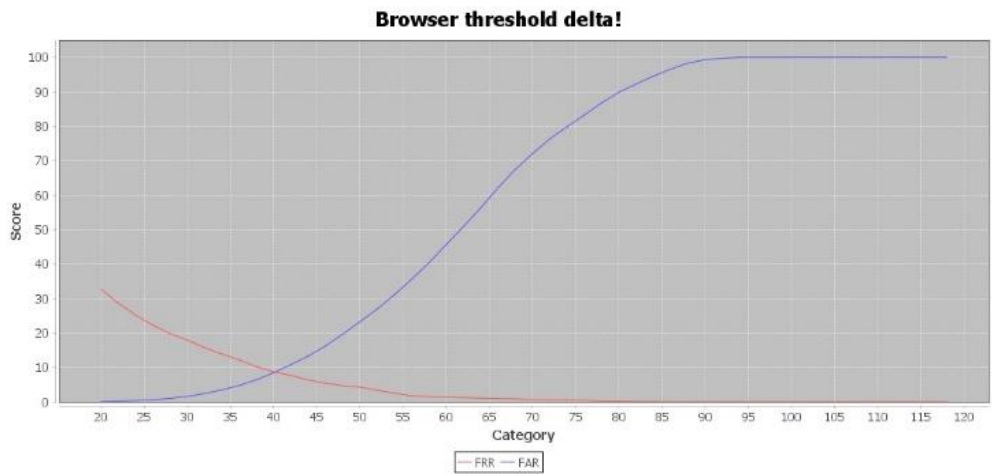

**Fig. 6.** FAR and FRR with our hybrid scheme.

## 5      Conclusions

In this paper, we have presented an efficient and innovative hybrid biometric template protection scheme which takes benefits of random orthonormal projection and fuzzy commitment while preventing their limitations. The hybrid scheme satisfies all requirements of: cancelability (revocability and diversity), security, and good performance. By using the lightweight approach for generating orthonormal matrix instead of the traditional Gram-Schmidt in random orthonormal projection, the computational complexity is reduced significantly. The experimental result also demonstrates that the accuracy of recognition is around 91% which stays the same in comparison with the case of none security method applied. That makes the proposal feasible.

## Acknowledgments

This research is funded by Vietnam National University - Ho Chi Minh City (VNUHCM) under grant number C2018-20-13. We also want to show a great appreciation to each member of DSTAR Lab for their enthusiastic supports and helpful advices during the time we have carried out this research.